\documentclass[12pt,a4paper,twoside]{article}
\usepackage{vlad_iaea2012_mod,color}
\usepackage{graphicx}
\usepackage{wrapfig}
\usepackage{amssymb}

\paper{Newton-Raphson method for transcendental equations in complex space}{I. CHAVDAROVSKI}	
\begin{document}

\title{COMPLEX SPACE NEWTON-RAPHSON SOLUTION OF THE DISPERSION \\
RELATION OF LOW FREQUENCY ALFV\'EN WAVES IN FUSION PLASMAS
}\\

\author{I. CHAVDAROVSKI\\
Korea Institute of Fusion Energy,
 169-148 Gwahak-ro, Yuseong-gu,
Daejeon, 34133, Republic of Korea \\
{Email: chavdarovski@gmail.com}}

\linespread{1} \small \normalsize
\begin{abstract}

\hspace{1cm}
Newton-Raphson method  in complex space is used to numerically solve the transcendental dispersion
relation of low frequency shear Alfv\'en waves in fusion plasmas. The equation exhibits several physically
relevant basins of attraction. The method is shown to be efficient and fast in solving this equation,
with very small portion of non-converging points.
\end{abstract}

\linespread{1.} \small \normalsize

\section{INTRODUCTION}
\label{sec:intro}

Newton-Raphson method is an algorithm for finding the roots of an equation $f(x)=0$, derived from the Taylor expansion of the function in the vicinity of a root. The method has been used for solving non-linear equations with single or multiple variables. If $f: R\rightarrow  R$ is a continuous and differentiable function of one argument $x \in R$, then the iterations leading to a solution of $f(x)=0$ is given by
\begin{eqnarray}\label{e:iter}
x_{n+1}=x_{n}-\frac{f(x_{n})}{f'(x_{n})}\,\,,
\end{eqnarray}
where the starting point $x_0$ is in the vicinity of the assumed root and $n$ is the iteration count. The calculation stops when $f(x_n)<\epsilon$, where $\epsilon\,\in R$ is the designated accuracy of the problem.
The method is also applicable in complex space and vector space~\cite{Henrici}. If the initial value $x_0$ is appropriately chosen to be near an expected root the convergence of the iterations is quadratic, whereas poor choice of $x_0$ can lead the iterations away from the roots. There are several situations in which the Newton-Raphson method can fail. If $f$ or $f'$ are not continuous, the solution can overshot and diverge. Further, Eq.~(\ref{e:iter}) has singularities at the local minima and maxima ($f'=0$). The method can also fail if the iterations enter a infinite cycle where $x_n$ oscillates between two of more values.

There are several cases in which the convergence is too slow, one of which is when the multiplicity of the zero of the function is higher than $1$, i.e. $f(x)\propto (x-a)^m$, where $m\in N,\, m>1$ and $a$ is one of the roots of the function. In this case the modified Newton's method $x_{n+1}=x_{n}-m f(x_{n})/f'(x_{n})$ is still quadratic. If $f'(a)=0$ then the convergence of the iterations is linear instead of quadratic. From Eq.(\ref{e:iter}) we see that the solution will also converge slowly if the derivative is too large ($f'\gg f$ ), making $x_{n+1}\approx x_n$. The transcendental equation solved in section \ref{sec:appli}$3$ has this type of regions of non-convergence.

In section \ref{sec:main}$2$ the main features and weaknesses  of the method are discussed. In section \ref{sec:appli}$3$ the transcendental complex dispersion relation of low frequency shear Alfv\'en waves in fusion plasmas is solved numerically and the basins of attraction are discussed. Concluding remarks are given in section \ref{sec:con}$4$.

\section{MAIN FEATURES}
\label{sec:main}

Expanding the function $f(x)$ in Taylor series around $x_n\approx a\,$, where $f(a)=0$ we have
\begin{eqnarray}\label{e:taylor}
f(a)=f(x_n)+f'(x_n)(a-x_n)+1/2 f''(\xi_n)(a-x_n)^2\,\,,
\end{eqnarray}
with $\xi_n$ being between $x_n$ and $a$. Dividing by $f'(x_n)$ we get that the error $\varepsilon_n=x_n-a$ changes as $\varepsilon_{n+1}= -\frac{f''(\xi_n)}{2f'(x_n)}\varepsilon_n^2\,$, i.e. Newton-Raphson method is quadratic.

\begin{figure}[h!]
\begin{center}
\includegraphics[width=0.75\textwidth]{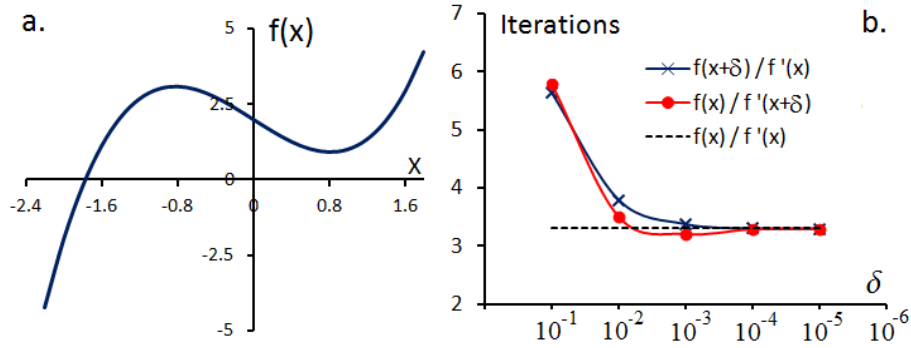}
\caption{a. Function $f(x)=x^3-2x+2$; b. Average number of iterations to solution around the root.}\label{fig:converge1}
\end{center}
\end{figure}

The problem of cyclic behavior mentioned above can be mitigated by adding a small 'error' $\delta$ in the calculation in order to 'escape' the infinite cycle by using
\begin{eqnarray}\label{e:assym1}
x_{n+1}=x_{n}-\frac{f(x_{n}+\delta)}{f'(x_{n})}\,\,.
\end{eqnarray}
Starting from a point near the solution, from Eq.(\ref{e:taylor}) we get:
$$\frac{f(x_n+\delta)}{f'(x_n)}+(a-x_n)=\frac{f(x_n+\delta)}{f'(x_n)}-\frac{f(x_n)}{f'(x_n)}-\frac{f''(\xi_n)(a-x_n)^2}{2 f'(x_n)}
\Rightarrow$$
$$a-x_{n+1}=\frac{f(x_n+\delta)-f(x_n)}{f'(x_n)}-\frac{f''(\xi_n)(a-x_n)^2}{2 f'(x_n)}\Rightarrow$$
\begin{eqnarray}\label{e:cyclic1}
\hskip -2em a-x_{n+1}=\delta+O(\delta^2)-\frac{f''(\xi_n)(a-x_n)^2}{2 f'(x_n)}\, \Rightarrow \varepsilon_{n+1}=-\frac{f''(\xi_n)}{2 f'(x_n)}\,\varepsilon_n^2+O(\delta^2).
\end{eqnarray}
In the last line we used the fact that the iterations stop when $f(x_n+\delta)=0$, hence the actual solution to the problem is $a=x_n+\delta$. The algorithm is still quadratic as long as $\delta^2$ is smaller than the acceptable error.
Similarly to this, we can add a small 'error' by calculating the derivative at $x_n+\delta$, obtaining
\begin{eqnarray}\label{e:assym2}
x_{n+1}=x_{n}-\frac{f(x_{n})}{f'(x_{n}+\delta)}\,\,.
\end{eqnarray}
Near the solution the error can be estimated as
$$\frac{f(x_n)}{f'(x_n+\delta)}+(a-x_n)=\frac{f(x_n)}{f'(x_n+\delta)}-\frac{f(x_n)}{f'(x_n)}-\frac{f''(\xi_n)(a-x_n)^2}{2 f'(x_n)}
\Rightarrow$$
$$a-x_{n+1}=f(x_n)\left(\frac{1}{f'(x_n+\delta)}-\frac{1}{f'(x_n)}\right)-\frac{f''(\xi_n)(a-x_n)^2}{2 f'(x_n)}\Rightarrow$$
\begin{eqnarray}\label{e:cyclic2}
\hskip 4em \varepsilon_{n+1}\approx-\delta\, \frac{f(x_n)f''(x_n)}{f'(x_n)^2}-\frac{f''(\xi_n)\varepsilon_n^2}{2 f'(x_n)} .
\end{eqnarray}
Since, the ratio $f(x_n)f''(x_n)/f'(x_n)^2$ can have arbitrary values this error is of order $\delta$ and will only
decrease quadratically for very small $\delta$. Hence, for not so small values of $\delta$, Eq.(\ref{e:assym1})
converges faster than Eq.(\ref{e:assym2}). This can be illustrated by taking for example the polynomial function $f(x)=x^3-2x+2$ (shown in Fig.\ref{fig:converge1}a) which has one zero at $x=-1.76929$, while the
points $x=0$ and $x=1$ form a periodic cycle of iterations not leading to a solution. Fig.\ref{fig:converge1}b shows the average number of iterations to convergence for starting points $x_0 \in [-2.2,1.5]$ for $\delta=0$, i.e Eq.(\ref{e:iter}), as well as  Eq.(\ref{e:assym1}) and Eq.(\ref{e:assym2}) for different values of $\delta$. The addition of $\delta$ can slow down the convergence, but this change is negligible for small $\delta$, while the correction to Eq.(\ref{e:iter}) is useful when the iterations lead to a cycle.

\begin{figure}[h!]
\begin{center}
\includegraphics[width=0.8\textwidth]{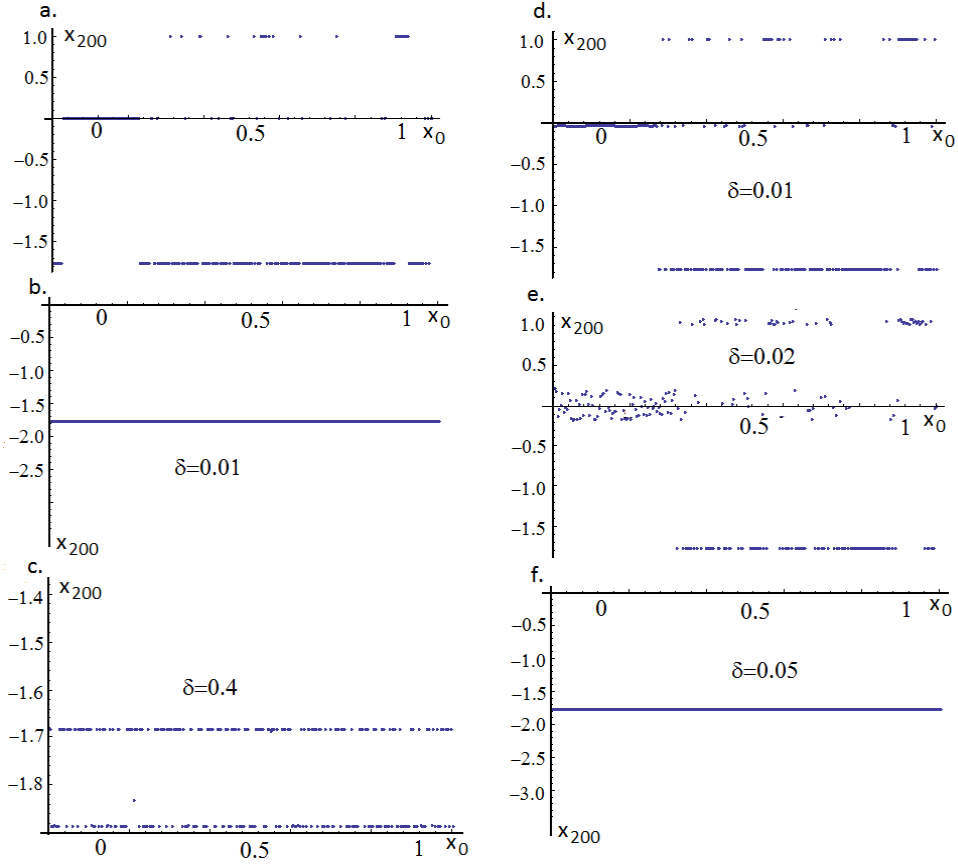}
\caption{a. Convergence of Eq.(\ref{e:iter}); b and c. Convergence of Eq.(\ref{e:assym1}); d,e and f.
Convergence of Eq.(\ref{e:assym2})}\label{fig:converge2}
\end{center}
\end{figure}

Regions of points in the near
vicinity of $x=0$ and $x=1$ can also form a cycle. In general there is a fine (fractal) structure,
where points very close to each other can lead to different solutions~\cite{Dence}.
In Fig.\ref{fig:converge2}a, we give $x_n$ after $n=200$ iterations for starting
points $x_0\in [-0.15,1.25]$, with a resolution of $\Delta x_0=0.01$. For the dots in the figure with $x_{200}\in\{0,1\}$ the iterations are in a cycle, while all the other starting points give the correct solution. Significant portion of this interval is
non-convergent. Applying Eq.(\ref{e:assym1}) and Eq.(\ref{e:assym2}) will not necessarily solve the cyclic behavior problem, since for small values of $\delta$, the iterations might find different points of oscillation near the original $x_{0}\in\{0,1\}$.
Comparing again Eq.(\ref{e:cyclic2}) and Eq.(\ref{e:cyclic1}) shows that using Eq.(\ref{e:assym2}) the iterations 'escape' the cycle more efficiently. This is demonstrated in Fig.\ref{fig:converge2}b, where for small $\delta=0.01$ Eq.(\ref{e:assym2}) gives the correct solution for all starting point, while for the same value (Fig.\ref{fig:converge2}d) has still too many cycles around $x=0$ and $x=1$. For larger $\delta=0.05$ the first method is also successful (Fig.\ref{fig:converge2}f). For large $\delta=0.4$ as expected in (Fig.\ref{fig:converge2}c) we see a degradation of the solutions, since the tangential premise of the method no longer applies.

Eq.~(\ref{e:iter}) is valid in complex space~\cite{Henrici} for a function $\Lambda: C \rightarrow C$ with a complex argument $\omega=\omega_r+i\omega_i \in  C\,$, where  $\omega_r,\omega_i \in  R$, assuming $\Lambda'=d\Lambda/d\omega$ and $d\Lambda'/d\omega$ are continuous functions in the vicinity of the roots. In the next section we will use the Newton-Raphson method to find the zeros of a complex transcendental function $\Lambda(\omega)=0$, for which the iterations are given by
\begin{eqnarray}\label{e:newton}
\omega_{n+1}=\omega_{n}-\frac{\Lambda(\omega_{n})}{\Lambda'(\omega_{n})}\,\,.
\end{eqnarray}
The method in complex space exhibits basins of attraction, which for many functions have fractal structure (see the next section), while for some of the starting points in the complex plane the iterations will diverge. This is the case with many of the iterative methods, since for example for polynomials of order $4$ or higher any iterative method will have open regions where the iterations diverge~\cite{McMullen}.
Additionally, for functions $\Lambda(\omega)$ containing $\sqrt[m]{\omega}$, where $m\in Z$, $\Lambda$ and $\Lambda'$ can have discontinuities where $arg(\omega)=(2k+1)\pi/ m $, with $k\in Z$.

The complex plane problem is similar to the $2$-dimensional vector space Newton's method. Since $\Lambda(\omega)=0 \Longleftrightarrow Re(\Lambda)=Im(\Lambda)=0$ and the variables $\omega_r$ and $\omega_i$ are independent, we are practically solving a system of two equations. We can pose the problem in $R^2$ vector space, where the argument is
$\vec{\omega}=(\omega_r,\omega_i)\in R^2$, while the function $\vec{{\bf{\Lambda}}}=(\Lambda_r,\Lambda_i) \in R^2$, where we used the notation $Re(\Lambda)=\Lambda_r$, $Im(\Lambda)=\Lambda_i$. Here, the notation $\vec{\omega},\vec{{\bf{\Lambda}}}\in R^2$ for the vectors are distinct from the complex numbers $\omega, \Lambda \in C$.
The system of equations then becomes $\Lambda_r(\vec{\omega}_{n})=0$ and $\Lambda_i(\vec{\omega}_{n})=0$, which can be iteratively solved as
\begin{eqnarray}\label{e:vector}
\vec{\omega}_{n+1}=\vec{\omega}_{n}-J^{-1} \vec{\Lambda}(\vec{\omega}_{n})\,\,,
\end{eqnarray}
where the Jacobean matrix is given by
\[ J= \left( \begin{array}{cc}
\frac{\partial \Lambda_r}{\partial \omega_r} & \frac{\partial \Lambda_r}{\partial \omega_i}  \\
\frac{\partial \Lambda_i}{\partial \omega_r} & \frac{\partial \Lambda_i}{\partial \omega_i} \end{array}\right).\]

Recalling, $\partial \Lambda_r/\partial \omega_r=Re(\partial \Lambda/\partial \omega_r)$ and $\partial \Lambda_i/\partial \omega_r=Im(\partial \Lambda/\partial \omega_r)$, with $\partial/\partial \omega_r=\partial/\partial \omega$
and $\partial/\partial \omega_i=i \partial/\partial \omega$, we obtain
\[ J= \left( \begin{array}{cc}
Re(\frac{\partial\Lambda}{\partial\omega}) & -Im(\frac{\partial\Lambda}{\partial\omega})  \\
Im (\frac{\partial\Lambda}{\partial\omega}) & Re(\frac{\partial\Lambda}{\partial\omega}) \end{array}\right)\]
and the Jacobian determinant is $|J|=|\partial\Lambda/\partial\omega|^2$, while the inverse matrix becomes
\[ J^{-1}= \frac{1}{|J|}\left( \begin{array}{cc}
Re(\frac{\partial\Lambda}{\partial\omega}) & Im(\frac{\partial\Lambda}{\partial\omega})  \\
-Im (\frac{\partial\Lambda}{\partial\omega}) & Re(\frac{\partial\Lambda}{\partial\omega}) \end{array}\right).\]
Eq.(\ref{e:vector}) gives two simultaneous iterations
\begin{eqnarray}\label{e:2}
\omega_{r,n+1}=\omega_{r,n}-\frac{1}{|J|} \left(Re(\frac{\partial\Lambda}{\partial\omega}) \Lambda_r+Im(\frac{\partial\Lambda}{\partial\omega})\Lambda_i\right)\,\,,
\end{eqnarray}
and
\begin{eqnarray}\label{e:3}
\omega_{i,n+1}=\omega_{i,n}-\frac{1}{|J|} \left(-Im(\frac{\partial\Lambda}{\partial\omega}) \Lambda_r+Re(\frac{\partial\Lambda}{\partial\omega})\Lambda_i\right)\,\,.
\end{eqnarray}
Multiplying Eq.(\ref{e:3}) by $i$ and adding it to Eq.(\ref{e:2}) yields
\begin{eqnarray}\label{e:final}
\hskip -2em \omega_{n+1}&=&\omega_{n}-\frac{1}{|\partial\Lambda/\partial\omega|^2} \left(Re(\frac{\partial\Lambda}{\partial\omega}) \Lambda_r+Im(\frac{\partial\Lambda}{\partial\omega})\Lambda_i-Im(\frac{\partial\Lambda}{\partial\omega}) i\Lambda_r+Re(\frac{\partial\Lambda}{\partial\omega})i\Lambda_i\right) \nonumber  \\ &=& \omega_{n}-\frac{1}{|\partial\Lambda/\partial\omega|^2} \Lambda \left(Re(\frac{\partial\Lambda}{\partial\omega}) -i Im(\frac{\partial\Lambda}{\partial\omega})\right) \nonumber  \\ &=& \omega_{n}-\frac{\Lambda}{Re(\frac{\partial\Lambda}{\partial\omega}) +i Im (\frac{\partial\Lambda}{\partial\omega})} = \omega_{n}-\frac{\Lambda}{\frac{\partial\Lambda}{\partial\omega}}\,\,,
\end{eqnarray}
which is exactly Eq.(\ref{e:newton}), showing that the two problems are identical. The vector method fails if $J$ is a singular matrix, which in single variable problem corresponds to $f'(x_n)=0$, while in complex space $\Lambda'(\omega_n)=0$. The vector solution is valid in general case in $R^m$ vector space, for $m\in N$, provided that $J^{-1} \neq 0$.

Equivalent problem to $\Lambda(\omega)=0$ in complex space is finding the minimum of $|\Lambda(\omega)|$, which also can use the Newton-Raphson as an optimization method. This solution has the same failures previously mentioned, with $|\Lambda|'=0$ corresponding to finding the local minimum of $|\Lambda|$ where $\partial |\Lambda|/\partial \omega_r=\partial |\Lambda|/\partial\omega_i=0$. Unlike, for example the steepest descent method that will almost certainly find
the local minimum, Newton-Raphson method is efficient in avoiding it and finding the root at $|\Lambda(\omega)|=0$, instead.
For that reason, this method is appropriate for functions that have multiple minima, like many transcendental functions do.

\section{Example of a complex transcendental equation}
\label{sec:appli}

The harmonic oscillations of the perturbed quantities of the plasma, like density, electric and magnetic field, have a form $\propto e^{-i \omega t }=e^{\omega_i t}\,e^{-i\omega_r t}$, where the real part $\omega_r$ represents the oscillation frequency, while the imaginary $\omega_i$ is the growth (or damping) rate. In plasmas, Alfv\'en wave is an incompressible transverse wave
with a frequency $\omega=k_\parallel v_A$, where $k_\parallel$ is the wave number, while $v_A$ is the so-called Alfv\'en velocity~\cite{Alfven}.
The transcendental equation solved in this section is the dispersion relation of the so-called Low frequency shear Alfv\'en waves in tokamak plasmas, written as $i \Lambda(\omega) = \delta \bar{W}$~\cite{TsaiChen}, where
$\delta \bar{W} \in C$ represents the potential of resonant and non-resonant wave-particle interactions in fusion plasmas. Here, $\Lambda$ is the general plasma inertia given in Ref.~\cite{ZC96} as
\begin{eqnarray}\label{e:lambdacir}
\Lambda^2 = \frac{\omega}{\omega_A^2} \left (\omega- {\omega_{\ast p}} \right)  +q^2\frac{\omega_{Ti}}{\omega_A^2} \left[\left(\omega-\omega_{\ast n}\right) F(\omega / \omega_{Ti})- \omega_{\ast T} G(\omega / \omega_{Ti})-\frac{\bar{N}^2(\omega / \omega_{Ti})}{\bar{D}(\omega / \omega_{Ti})} \right],
\end{eqnarray}
where
\begin{eqnarray}
\hskip -4em F(x) & =& x \left( x^2 + 3/2 \right) +  \left( x^4 + x^2 + 1/2
\right) Z(x) \, ,\nonumber \\
\hskip -4em G(x) & =& x \left( x^4 + x^2 + 2 \right) +  \left( x^6 + x^4/2
+ x^2 + 3/4 \right) Z(x) \, ,\nonumber \\
\hskip -4em \bar{N}(x)& =& \left(\omega-\omega_{*n}\right)[x+(1/2+x^2)Z(x)]- \omega_{*T}[x(1/2+x^2)+(1/4+x^4)Z(x)]\,,\nonumber \\
\hskip -4em \bar{D}(x)& =&\omega_{Ti} \left( 1 + \frac{T_i}{T_e} \right)
+ \left(\omega - \omega_{*n} \right)  Z(x)-
\omega_{*T}
\left[ x + \left(  x^2 - 1/2 \right)  Z(x) \right]
\label{eq:fgnd}
\end{eqnarray}
with
\begin{equation}\label{e:z}
Z(x)=1/ \sqrt\pi \int_{-\infty}^{\infty} e^{-y^2}/(y-x) \,dy
\end{equation}
being the \emph{plasma dispersion function} defined with complex argument $x \in C$~\cite{FriedConte,Fried68}.
This function is closely related to the imaginary Error function as $Z(x)=i \sqrt{\pi}e^{-x^2}(1+erf(ix))$,
as well as the Faddeeva function $w(x)=e^{-x^2}erfc(-ix)$~\cite{Faddeeva}.
The dispersion function is differentiable with $Z'(x)=-2 (1+x Z(x))$.

There are several parameters in Eq.(\ref{e:lambdacir}) that come from the plasma characteristics for which we will not give the physical meaning, but only consider them as given constants.  The function contains the so-called ion diamagnetic frequencies $\omega_{*p}=\omega_{*n}+\omega_{*T}$, the ratio of temperature of thermal electrons and ions $\tau=T_e/T_i$, the frequency of the periodic circulating motion of the ions around the torus $\omega_{Ti}$ and the Alfv\'en frequency $\omega_A\gg\omega,\, \omega_{*p},\, \omega_{Ti}$.
For simplicity we normalize all the frequencies to $\omega_{Ti}$ and take $\omega_A=10$, $\omega_{*n}=0.2$, $\omega_{*T}=0.1$ and $\tau=1$, while for the so-called tokamak safety factor $q=1.5$.
The wave- particle resonance $\omega/\omega_{Ti}=1$ is contained in the
$Z$-function. The solution of $\Lambda(\omega)=0$ gives the so-called \emph{accumulation point} of the continuous spectrum, but the numerical solution to the full dispersion equation $\Lambda+i\delta \hat{W}=0$ is not very different.
For a detailed description of the physical meaning of the problem read Refs.~\cite{TsaiChen,ZC96,chavdar2009,chavdar2009b,Zonetal2010,chavdar2014,riri2014}.

\begin{figure}[h!]
\begin{center}
\includegraphics[width=0.6\textwidth]{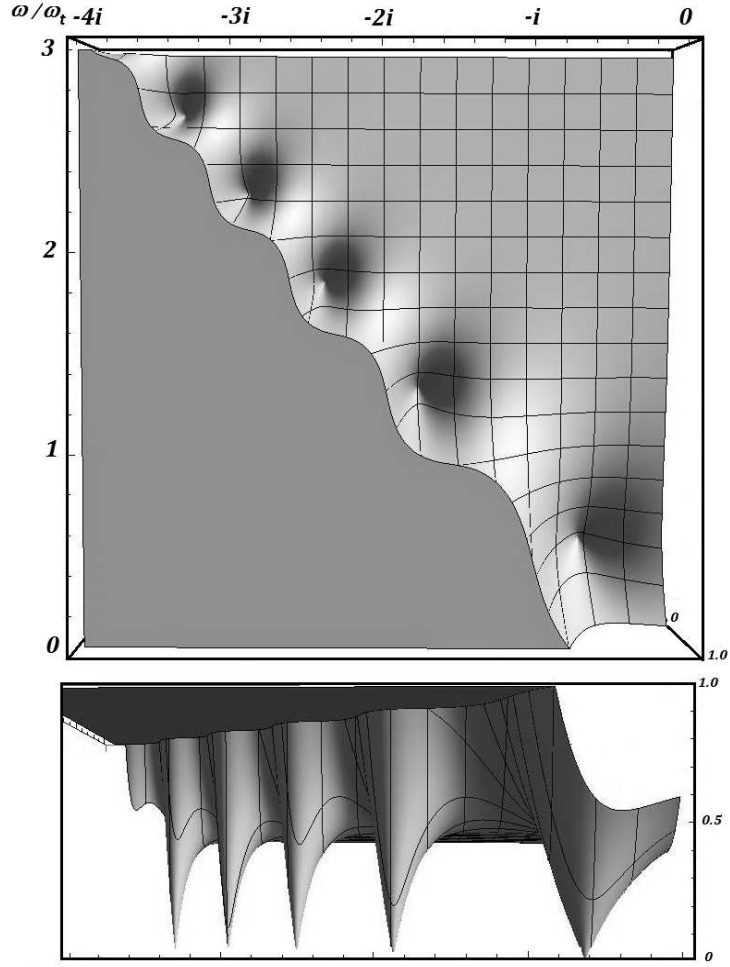}
\caption{$|\bar{D}|$ shown in the negative complex plane as function of normalized $\omega/\omega_{Ti}$.
The horizontal axis is the imaginary part of $\omega/\omega_{Ti}$.}\label{fig:D}
\end{center}
\end{figure}

In Fig.\ref{fig:D}, $|\bar{D}(x)|$ is given in the negative complex plane for the parameters mentioned above,
with a cut-off for large values.
We note here that this function can have many zeros in the complex plane, which give singularities in $\Lambda^2$.
However, the function is very steep around the roots of $|\bar{D}|=0$ (see Fig.\ref{fig:D}), so the iterations $x_n\equiv\omega_n$
would likely avoid these singularities.

The asymptotic limits of Eq.~(\ref{e:lambdacir}) can be determined using the expansions of the $Z(x)$ function in
the high ($x\rightarrow \infty\,,\omega\gg\omega_{Ti},\omega_{*p}$) and low frequency limits  ($x\rightarrow 0\,,\omega\ll\omega_{Ti}$).
When $x\rightarrow 0$~\cite{FriedConte,Fried68}
\begin{equation}\label{e:z0}
Z(x)=(-2x) [1-\frac{2}{3}x^2+\frac{4}{15} x^4+ \dots\,\,]+i\sqrt{\pi}e^{-x^2}\,\, .
\end{equation}
Keeping finite $\omega_{*n,T}$ the terms in Eq.(\ref{eq:fgnd}) become
\begin{eqnarray}
\bar{D}(x)\approx\omega_{Ti} \left( 1 + \frac{1}{\tau} \right)+\left(\omega-\omega_{*n}\right)(-2x^2 + i x \sqrt{\pi})- \omega_{*T}\,x [2x - i\sqrt{\pi}/2]\,,\nonumber
\end{eqnarray}
\begin{eqnarray}
\bar{N}(x) &=&
\left(\omega-\omega_{*n}\right)\frac{i\sqrt{\pi}}{2}-
\omega_{*T}\frac{i\sqrt{\pi}}{4}+{\mathcal{O}}(x^2)\,\,, \nonumber
\end{eqnarray}
\begin{eqnarray}
F(x) \approx x/2+ i\sqrt{\pi}/2 \, , \,\, G(x) \approx \frac{x}{2}+ \frac{3}{4}i\sqrt{\pi}
\, . \label{e:lowexpand}
\end{eqnarray}
The imaginary values that appeared in the equations above have no physical meaning, since they are consequence of going to frequencies way below the resonant frequency of the particles $\omega_{Ti}$. In Refs.~\cite{chavdar2009,chavdar2014} corrections were added to the functions in order to have the appropriate solutions at low frequency $\omega/\omega_{Ti}\ll 1$, which cancel out the imaginary parts of the functions. When this is taken into account, Eq.~(\ref{e:lambdacir}) has the low frequency limit~\cite{Rosenbluth}
\begin{eqnarray}\label{e:lowcirc}
\Lambda^2 = \frac{\omega}{\omega_A^2}\left( \omega- \omega_{\ast p}\right)
(1+0.5q^2)\,.
\end{eqnarray}
We notice here two obvious branches of solutions of $\Lambda=0$, the first one being
$\omega=0$ which represents the Magneto-hydrodynamic (MHD)~\cite{Alfven} description of Alfv\'en waves in plasmas, when the parallel wave number $k_\parallel=0$ . The second solution is around $\omega=\omega_{*p}$ which are the so-called Kinetic Ballooning Modes (KBM)~\cite{TsaiChen}. When the $\Lambda^2$ term is uncorrected like in Eq.(\ref{e:lambdacir}), the two solutions are away from the ones given in Eq.(\ref{e:lowcirc}). Here, for the sake of simplicity we will keep
the terms as they are given in Eq.(\ref{e:lowexpand}), still recalling the origin of the solutions is in the two branches $\omega=0$ and $\omega=\omega_{*p}$.

In the other asymptotic limit $x\rightarrow \infty$ ($\omega \gg \omega_T, \omega_{*p}$) we have~\cite{FriedConte,Fried68}
\begin{equation}\label{e:zinf}
Z(x)=\left(-\frac{1}{x} \right) [1+\frac{1}{2x^2}+\frac{3}{4 x^4}+\frac{15}{8 x^6}+\ldots\,\,]+i\sqrt{\pi}\sigma e^{-x^2}\,\, ,
\end{equation}
where
\[ \sigma = \left\{ \begin{array}{ll}
0, & Im(x) > 1/|Re(x)| \\
1, & Im(x) < 1/|Re(x)| \\
2, & Im(x) < -1/|Re(x)| \,,\end{array} \right.\]
and the terms become
$\bar{D}(x)\approx 1/\tau$, $\bar{N}(x)\approx-1$ and $F(x) \approx -7/(4x)$ giving
from Eq.~(\ref{e:lambdacir})
\begin{equation}\Lambda^2=\frac{\omega^2}{\omega_A^2}\left[1-q^2\frac{\omega_{Ti}^2}{\omega^2}\left(\frac{7}{4}+\tau\right)
\right]\, . \label{e:lambdahigh}
\end{equation}
This shows the third branch of solutions at $\omega=\omega_{BAE}$, where~\cite{ZC96}
\begin{equation}\label{e:omegaBAE}
\omega_{BAE}=q \omega_{Ti} \sqrt{\frac{7}{4}+\tau}\,\,
\end{equation}
is the frequency of the so-called Beta induced Alfv\'en eigenmodes (BAE)~\cite{Heidbrink,Turnbull}.
This solution is also approximate like the other two, while the exact solutions are dependent on the values of $\omega_{*n,T}$~\cite{ZC96,chavdar2014}.

Since $\Lambda^2=0  \Longleftrightarrow \Lambda=0$, Eq.(\ref{e:lambdacir}) shows it's more prudent
to solve the former equation, since applying the square root to $\Lambda^2$ would generate gaps in the function where the Riemann surface unwinds.
We are only looking at the positive frequency range $\omega_r\geq 0$ with the ordering $\omega_{*p} \sim \omega_{Ti}$.

\begin{figure}[h!]
\begin{center}
\includegraphics[width=0.8\textwidth]{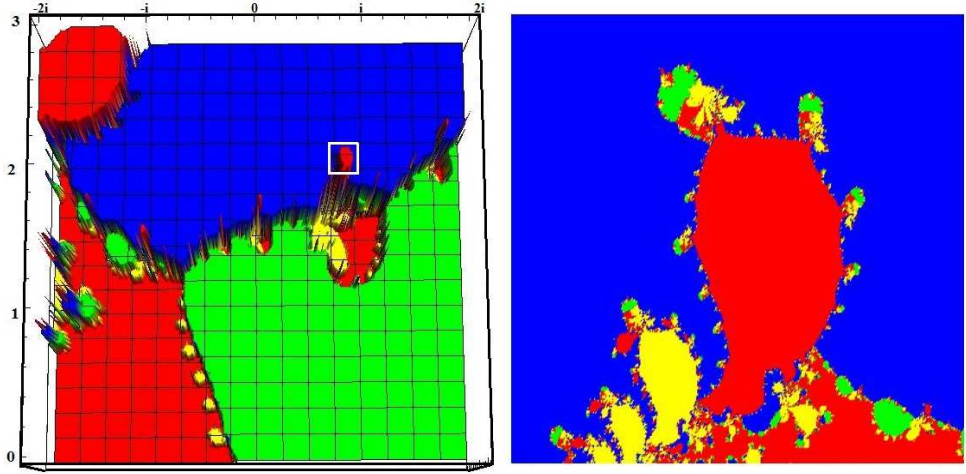}
\caption{(Left) The basins of attraction of $\Lambda^2=0$. (Right) Fractal structure of the basins.}\label{fig:fractal}
\end{center}
\end{figure}
Fig.(\ref{fig:fractal}) shows five different basins of attraction: the first one with red color in the lower left corner is the set of starting points leading to $\omega=0.138 - 0.702i$, which corresponds to the approximate solution $\omega=0$. The second branch with green color leading to $\omega=0.334 + 0.012i$ is the mentioned KBM, approximately given by $\omega\approx \omega_{*p}=0.3$.
The blue region is for the BAE branch $\omega=2.825 - 0.032 i$, where the analytical solution was $\omega_{BAE}=q \omega_{Ti} \sqrt{7/4+\tau}=2.487$. The solution in the upper left corner is from the BAE branch, but with large negative imaginary frequency $\omega_{BAE2}=2.772 - 1.479 i$. In the imaginary half-plane an
infinite number of points of attraction related to each branch can be found, with large negative imaginary values, which is typical for transcendental equations. In physical sense the waves with large negative imaginary frequency are not important since they are heavily damped ($e^{\omega_i t}\rightarrow 0$).

\begin{figure}[h!]
\begin{center}
\includegraphics[width=0.5\textwidth]{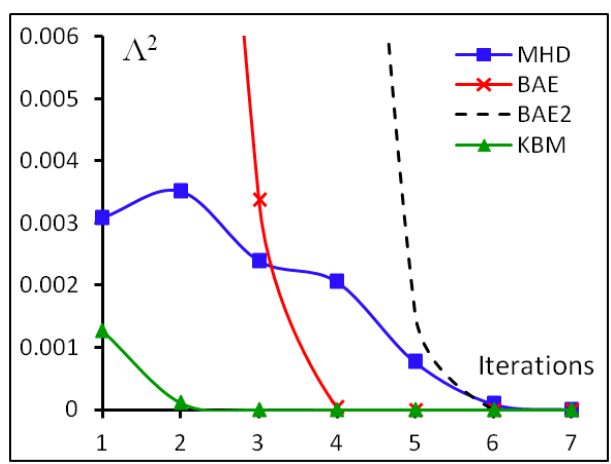}
\caption{Convergence of four different basins of solutions from Fig.(\ref{fig:fractal}).}\label{fig:converge}
\end{center}
\end{figure}

The yellow space in Fig.(\ref{fig:fractal}) are starting points that either converge to a solution with $\omega_r<0$ or don't converge after $n=100$ iterations. The portion of starting points that don't converge is negligible, and can be estimated to be about $0.00079$.
Closer look at these regions shows that most of the diverging paths come from the points where the derivative of $\Lambda^2$ is very big, yielding $\omega_{n+1}\approx\omega_n$ at frequencies where $\Lambda^2(\omega) \neq 0$. Additionally, small portion of the points lead to non-converging cycles. In both cases, the non-convergent space can be reduced using $$\omega_{n+1}=\omega_n-\frac{\Lambda^2(\omega_n)}{\frac{\partial (\Lambda^2)}{\partial \omega}|_{\omega_n+\delta}}\,.$$ In our case $\delta=10^{-4}(1+i)$ reduces the non-convergent space by $10\%$, while large $\delta=0.1 (1+i)$ as expected degrades the solution.
In general, there is a conflict between the need to reduce the non-convergent space and the slowing down of the original algorithm. Since large portion of the points here are regular (Fatou set), it is advisable to start the iterations from a different point, when a threshold of iterations without reaching a solution is passed.
Fig.(\ref{fig:converge}) shows the convergence for a starting point around each of the four mentioned attractors. The convergence for the regular points is very fast, quadratic and usually takes less than $7$ iterations to arrive below the acceptable error, here taken to be $|\Lambda^2|<10^{-6}$.

\section{Conclusion}
\label{sec:con}

We have used the Newton-Raphson method in complex space to solve the transcendental equation of low frequency shear Alfv\'en waves in fusion plasmas containing polynomials of order higher than four. Up to five basins of attractions are identified in the observed frequency range, three of which are physically relevant branches for this work. However, each branch is expected to generate infinite number of attractors in the negative imaginary space. Due to the transcendental nature of the equation, small portion of non-converging points can be found at locations where the function is too steep, thus slowing down the calculation. Since the convergence from the regular points is very fast, rather than modifying the method, it is advisable to start the iteration process from a new starting point if the acceptable threshold of iteration is passed without convergence.

ACKNOWLEDGMENT

This work is supported by R\&D Program through the Korean Institute of Fusion Energy (KFE)
funded by the Ministry of Science and ICT of the Republic of Korea (KFE-EN2141-7).

\end{document}